\documentclass[pdflatex,sn-mathphys-num]{sn-jnl}

\usepackage{graphicx}%
\usepackage{multirow}%
\usepackage{amsmath,amssymb,amsfonts}%
\usepackage{amsthm}%
\usepackage{mathrsfs}%
\usepackage[title]{appendix}%
\usepackage{xcolor}%
\usepackage{textcomp}%
\usepackage{manyfoot}%
\usepackage{booktabs}%
\usepackage{algorithm}%
\usepackage{algorithmicx}%
\usepackage{algpseudocode}%
\usepackage{listings}%

\theoremstyle{thmstyleone}%
\theoremstyle{thmstyletwo}%

\theoremstyle{thmstylethree}%

\begin{document}

\title[Reducing Sexual Predation and Victimization Through Warnings and Awareness among High-Risk Users]{Reducing Sexual Predation and Victimization Through Warnings and Awareness among High-Risk Users}


\author*[1,3,a]{\fnm{Masanori} \sur{Takano}}\email{takano\_masanori@cyberagent.co.jp}

\author[2,b]{\fnm{Mao} \sur{Nishiguchi}}\email{nishiguchi@crimson.q.t.u-tokyo.ac.jp}

\author[2,c]{\fnm{Fujio} \sur{Toriumi}}\email{tori@sys.t.u-tokyo.ac.jp}

\affil*[1]{\orgdiv{Multidisciplinary Information Science Center}, \orgname{CyberAgent, Inc.}, \orgaddress{\city{Tokyo}, \country{Japan}}}

\affil[2]{\orgdiv{School of Engineering}, \orgname{The University of Tokyo}, \orgaddress{\city{Tokyo}, \country{Japan}}}

\affil[3]{\orgdiv{Center of Advanced Research for Human-AI Symbiosis Society}, \orgname{Keio University}, \orgaddress{\city{Tokyo}, \country{Japan}}}

\affil[a]{\url{https://scholar.google.co.jp/citations?user=ATyEdMsAAAAJ}}
\affil[b]{\url{https://scholar.google.co.jp/citations?user=FKw6S60AAAAJ}}
\affil[c]{\url{https://scholar.google.co.jp/citations?user=RzN_HF8AAAAJ}}

\abstract{Online sexual predators target children by building trust, creating dependency, and arranging meetings for sexual purposes. This poses a significant challenge for online communication platforms that strive to monitor and remove such content and terminate predators' accounts. However, these platforms can only take such actions if sexual predators explicitly violate the terms of service, not during the initial stages of relationship-building. This study designed and evaluated a strategy to prevent sexual predation and victimization by delivering warnings and raising awareness among high-risk individuals based on the routine activity theory in criminal psychology. We identified high-risk users as those with a high probability of committing or being subjected to violations, using a machine learning model that analyzed social networks and monitoring data from the platform. We conducted a randomized controlled trial on a Japanese avatar-based communication application, Pigg Party. High-risk players in the intervention group received warnings and awareness-building messages, while those in the control group did not receive the messages, regardless of their risk level. The trial involved  12,842 high-risk players in the intervention group and 12,844 in the control group for 138 days. The intervention successfully reduced violations and being violated among women for 12 weeks, although the impact on men was limited. These findings contribute to efforts to combat online sexual abuse and advance understanding of criminal psychology.}

\keywords{Online sexual grooming, Graph neural network, Randomized controlled trial}



\maketitle

\section{Introduction}
Online sexual victimization, particularly among children, is increasing every year~\citep{Ringenberg2022,Subramanyam2024}.
Platforms are responsible for providing a safe online environment in a society where individuals cannot avoid using the internet. Regarding services, several platforms have banned sexual predation, such as obscene messages, sexual harassment, sexual grooming, and requesting or uploading sexual images. They monitor sexual predators, remove their content, and terminate predators' accounts.

However, sexual predators can evade the monitoring~\citep{Lykousas2018,Lykousas2021,Ringenberg2022,Borj2023}. This is because several of their sexual grooming posts are not immediately considered violative of the terms of service~\citep{Wolak2008,Ringenberg2022}. First, they target children who dream of adult romance, are interested in sexuality, live nearby, and have no connections with the police or lawyers.
Second, they build trust with children, isolate them from their parents and friends, make the children dependent on themselves, and gradually make them conscious of sexual relationships.
Finally, they are baited for sexual purposes.
Explicit violations arise only in the final step.

Previous studies have captured signs of sexual grooming in these first and second steps~\citep{Cano2014,Lykousas2018,Nishiguchi2024,Subramanyam2024}.
In capturing sexual grooming, behavioral logs of seeking targets and establishing relationships are effective.
This is because predators' communication behavior cannot conceal the features of grooming, although they can avoid posting direct sexual expressions.
Some studies using networking data have developed models to detect sexual predators before the final step, for example~\citep{Lykousas2018,Yokotani2021_chb2,Nishiguchi2024}.

\begin{figure}[t!]
  \begin{center}
    \includegraphics[width=0.8\columnwidth]{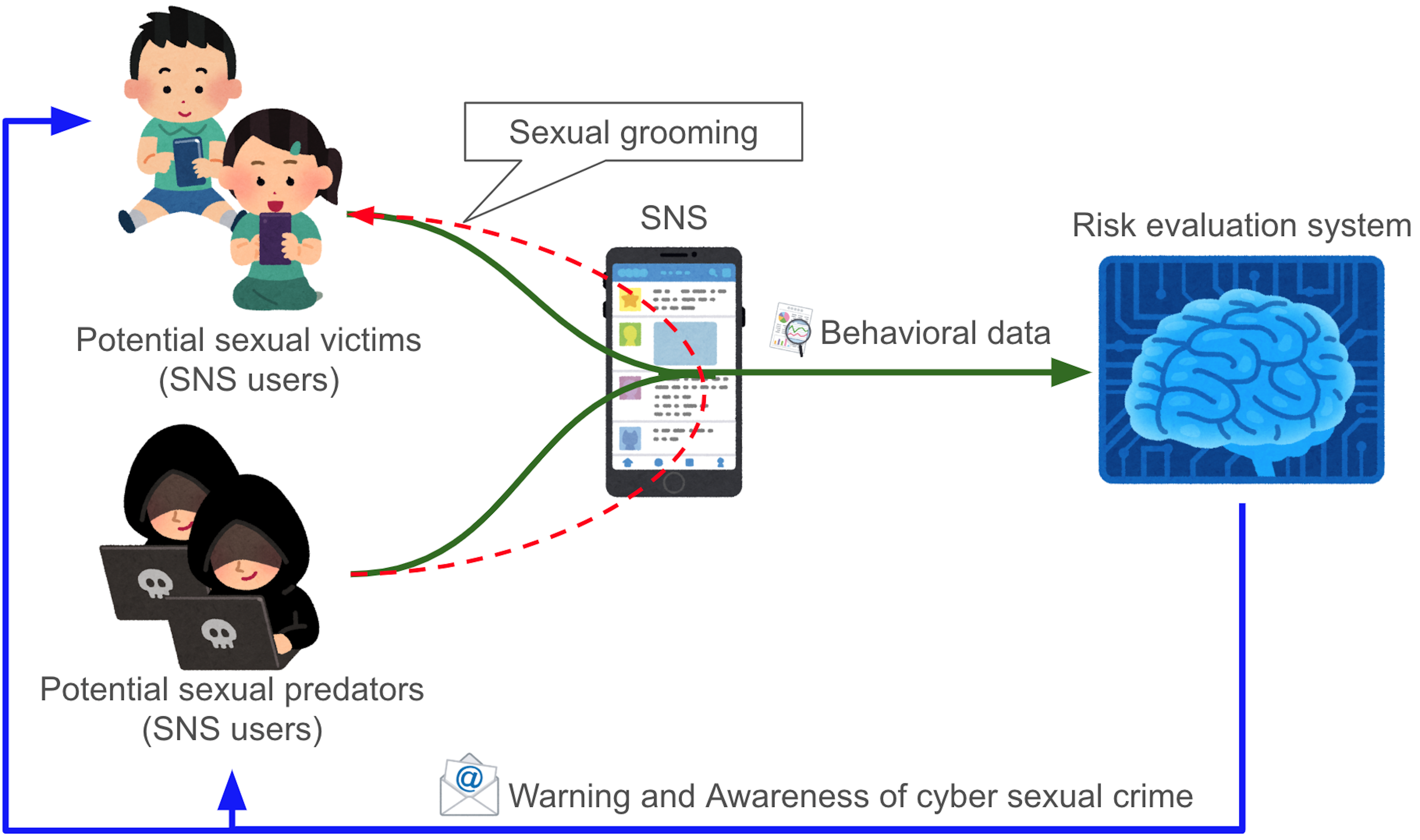}
    \caption{Schematic diagram of the experiment}
    \label{fig_ponchi}
  \end{center}
\end{figure}

However, there is a practical issue regarding the pre-detection of sexual predators.
Platforms are limited in their actions to pre-detected sexual predators because the predators' behavior may not yet constitute a violation.
Only the last step could be explicitly addressed on these platforms.
An unclear ban is against the predators, users, as well as the society\footnote{\url{https://www.politico.com/news/2022/12/15/twitter-suspends-journalists-musk-00074261},\url{https://www.newarab.com/news/x-suspends-hundreds-palestinian-accounts-amid-gaza-war}}.

Under these circumstances, warnings and awareness-building for predators can effectively inhibit violations and crimes.
Criminals with an extensive offending history (prolific offenders) do not tend to offend if they are visited homes and builded awareness by the police~\citep{Ariel2019}.
A comprehensive campaign related to online child grooming by the UK's Lucy Faithfull Foundation (LFF) facilitated access to helplines for potential offenders. 
This campaign discouraged them from engaging in online child sexual abuse~\citep{Walsh2023}.
In this campaign, LFF informed about the illicitness of child sexual abuse, damage caused to children, and the negative consequences for offenders after arrest.
Facilitating reconsidering posting violation comments effectively discouraging violation posts~\citep{Katsaros2022}. 
This approach has been adopted by many platforms on public comment forums\footnote{For example, \url{https://about.fb.com/news/2019/12/our-progress-on-leading-the-fight-against-online-bullying/}, 
\url{https://blog.youtube/news-and-events/make-youtube-more-inclusive-platform/}, 
\url{https://medium.com/jigsaw/helping-authors-understand-toxicity-one-comment-at-a-time-f8b43824cf41}, 
\url{https://newsroom.pinterest.com/en/creatorcode}, 
and \url{https://ameblo.jp/staff/entry-12612189833.html.}}.

Here, we consider inhibiting sexual predation and victimization by delivering warnings and awareness building to high-risk individuals (potential sexual predators and victims; Fig.~\ref{fig_ponchi}).
This is because administrators can send warnings and awareness-building messages even if individuals do not violate the terms of service.
The purpose of this messaging is as follows: 1) to inform potential victims that their behavior is high-risk and that the application administrator is addressing it; and 2) to reiterate to potential predators the prohibited behavior and inform them that the operation is monitoring their behavior as a measure to prevent violations.
Interventions using such messaging can change people's behavior, for example, by decreasing hate speech~\citep{Munger2017} and conducting energy conservation~\citep{ito2018}.
These messaging interventions have shown that their effects can be reasonably long-lasting.

We sent warning and awareness-building messages to two types of high-risk users: those at risk of violation and those at risk of being violated.
This is because the sexual violations of the terms of service do not always correspond to sexual grooming.
For example, if a child manipulated by a sexual predator sends the child sexual photos to the predator. Here, the child becomes both a violator and a victim.
For the same reason, we did not correspond potential predators/victims to violators/the being violated users in our data analyses.

\begin{figure}[t]
\centering
  \includegraphics[width=0.6\columnwidth]{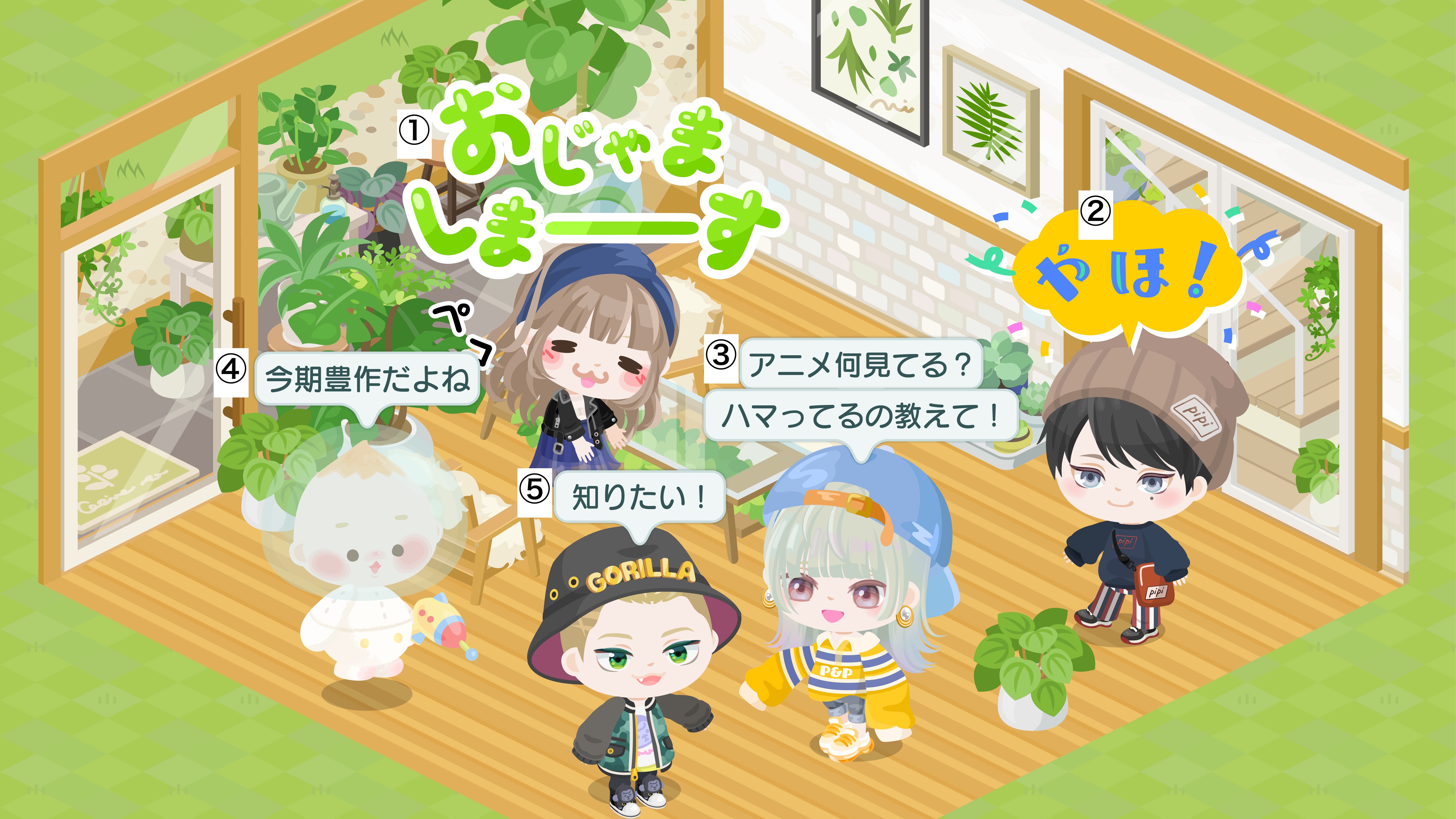}
  \caption{
  Players chat via their avatars in Pigg Party. Translation of this figure: ``Mind if I join in?,’’ 2: ``Hey,’’ 3: ``Which anime are you watching? Tell me what you're hooked on!,’’ 4: ``There are so many good ones this season, aren't there?,’’ and 5: ``I want to know!.’’
  }
  ~\label{fig_piggparty}
\end{figure}

We tested this approach using a popular Japanese avatar communication service, ``Pigg Party.’’
Pigg Party players are typically young women~\citep{MasanoriTakano2019}. 
As there are several potential targets for sexual predators, child protection is significantly essential.

The contributions of this study are as follows: 
{\bf 1)} We proposed a simple approach (sending warnings and awareness-building messages), which can be introduced to various communication platforms for decreasing online sexual predation and victimization.
{\bf 2)} We evaluated the effectiveness of this approach on an existing platform. The approach reduced the number of women violations and those being violated relating to sexual grooming.
This effect was long-term (12 weeks).
However, the impact of this approach on men was limited.

\section{Theoretical Rationale and Hypotheses}

The theoretical rationale of our approach is based on the routine activity theory~\citep{Lawrence1979} in criminal psychology.
This theory explains crime incidents when ``a likely offender'' encounters ``a suitable target'' ``the absence of a capable guardian.''
On online communication platforms, the role of an administrator, as a guardian who monitors users, is crucial.
This implies that we can inhibit violations by creating the impression of administrative oversight.
For example, in the physical world, closed-circuit television (CCTV) surveillance cameras suppress the occurrences of crimes~\citep{Piza2019}.
In an avatar communication service, observing the termination of a friend's account for violations can inhibit a user's own violation by making them realize the platform's monitoring~\citep{yokotani_tele2022}.
The routine activity theory states that crime opportunities increase when targets visit dangerous locations~\citep{Hindelang1978}.
Therefore, the inhibition of visiting dangerous places by targets should be reduced, potentially decreasing the likelihood of them being violated.
In Pigg Party, a previous study~\citep{Yokotani2021_chb2} showed that online violations occur between 8 P.M. and 5 A.M., particularly during midnight.
This finding is consistent with earlier studies conducted on other platforms~\citep{Parapar2014,Rodrigues2016}.
Thus, players who want to avoid being violated should refrain from using these platforms at night.

Thus, we propose the following hypotheses:
\begin{itemize}
    \item {\bf H1}: Sending warnings and awareness-building messages decreases violations.
    \item {\bf H2}: Sending the messages also decreases the likelihood of being violated.
\item {\bf H3}: When H2 is true, players receiving the messages avoid using such platforms at night.

\end{itemize}

The impacts of our proposed approach might decline over time. 
This is because, over time, people may forget awareness. After all, the learning effect decreases exponentially (Ebbinghaus forgetting curve~{Ebbinghaus1885,Murre2015}).
Repeated messaging may help them to remain vigilant. 
However, repeated messaging without a strategy might lose its impact over time~\citep{ito2018}.
Habituation theory~\citep{Thompson1966,Groves1970,Rankin2009} predicts that the effect of repeated messaging could decrease over time.
This is because repeated stimulus presentation might cause a decrease in reaction to the stimulus. 

Therefore, the impact of warnings and awareness-building messages diminishes as time passes after receipt, and individuals forget or get accustomed to receiving messages repeatedly.
Thus, we propose the following hypotheses:
\begin{itemize}
    \item {\bf H4}: Violation inhibition effect decreases as time passes after receiving the message.
    \item {\bf H5}: Avoiding being violated effect decreases as time passes after receiving the message.
\end{itemize}
Conversely, some intervention experiments have shown that the effects can be reasonably long-lasting, for example~\citep{Munger2017,ito2018}.

\section{Methods}

\subsection{Testing Platform: Pigg Party}

Pigg Party is a popular Japanese avatar communication application\footnote{A previous study reported that there were at least 550,000 active players during half a year~\citep{Yokotani2021_chb2} developed by CyberAgent, Inc.
The study excluded players who chat with a specific user more than 100 times daily from their analysis.}.
In Pigg Party, players usually communicate using personalized avatars in virtual spaces (Fig.~\ref{fig_piggparty}).
Pigg Party players were predominantly young women with a female ratio of 61\% and teenager ratio of 65\%~\citep{MasanoriTakano2019}.
Given the high number of potential targets for sexual predators, child protection becomes critically important.
The Pigg Party administrator has created and disseminated an awareness page for child grooming, targeting all players\footnote{\url{https://lp.pigg-party.com/rules}}.

Our dataset did not include player age and gender information.
We used avatars' age and gender as approximations for players' ages and genders because those of players and avatars indicated high similarities based on a previous survey in Pigg Party\footnote{The researchers of the previous survey have provided this data to us. After a review, we will describe the survey information.}.
A total of 87.1 \% of male players used male avatars, and 93.5\% of female players used female avatars.
The correlation between avatars' age and players' age was 0.769.

This platform provides five types of social interactions.
We focused on violations in two main types of communication: direct messages (DM) and avatar chats (AC).
Players can use the direct message function to talk to their friends in a talk group through text messages, similar to other SNS messaging functions, such as Facebook Messenger and X Direct Message.
Players invite their friends to a talk group\footnote{In Pigg Party, two players can be ``friends'' by mutually accepting each other's requests.}.
This implies that these talk group members have at least one friend in the group.
The group size ranges from 2 to 15.
In avatar chats, players can communicate synchronously through their avatars in virtual spaces (rooms), as shown in Fig.~\ref{fig_piggparty}.
All players have their own rooms and can visit the other players' rooms. 
In addition to sending text messages, players can respond with dozens of avatar animations, known as avatar actions.
Players can talk to players in the same room through this communication method. 

We focused on violations related to sexual grooming, such as asking for and sharing personal information, asking for a date, requesting sexual photos, and making sexual remarks.
The Pigg Party administrator monitors players' behavior to ensure compliance with the terms of service\footnote{\url{https://lp.pigg-party.com/terms}}.
We used the results of the monitoring related to sexual grooming to learn a model and evaluate the intervention effect.

We used all types of social interaction logs, including DM and AC, to estimate the risk of violations and the likelihood of being violated.
Other three types of social interactions are as``follows'',  ``board posts and comments'',  and ``like.''
\begin{itemize}
\item {\bf Comment}: Pigg Party provides a board for each player, like Facebook's feed.
Communication on boards is asynchronous; all players can read all players' boards and add comments, except for the players are blocked by the board owner.

\item {\bf Follow}: Players follow other players. They can quickly check the public activities of players who they follow.
If two players follow each other, they are regarded as having a friendly relationship, which leads to sending a DM. 

\item {\bf Like}: Players send ``like'' by one click. This is similar to Facebook's ``like'' button. The application notifies a like sender to like receivers.
\end{itemize}

\subsection{Experimental Design}

This experiment was conducted by the Pigg Patty administrator in collaboration with the authors as a countermeasure against online sexual grooming. 
This study analyzes the data received from that experiment.

\subsubsection{Basic Principles}

The intervention experiment was designed through in-depth communication between the Pigg Party application provider and the authors to maintain a healthy relationship between the provider and their players.
Additionally, we made the message simple and specific so that children could easily understand it.

Owing to the nature of pre-violation risk assessment, players with a high risk of violation or being violated possess the following characteristics:
They have not yet actually acted in violation, even if their risk of violation is high. 
As mentioned above, they might be potential victims.
Similarly, individuals with a high risk of being violated have not yet been violated.
They might be potential sexual predators.

We should not brand high-risk violation players as sexual predators or high-risk being-violated players as victims.
The following intervention policies were implemented:
\begin{itemize}
    \item {\bf Intervention targets}: To minimize interference with players who pose no problems, interventions should be limited to a few high-risk individuals. 
    \item {\bf Warning and awareness building message}: We explicitly showed the basis for sending the message. Specifically, we stated that ``your behavior tendencies have been similar to those who have had trouble in the past.''
    Further, as mentioned above, there are cases where it is impossible to distinguish between sexual predators and victims from violation or being violated risks. Therefore, we used the same message content for high-risk users of violation and those being violated. 
    The content was primarily for awareness building. Hence, the warnings are indirect.
\end{itemize}

\subsubsection{Intervention Message}

According to the basic principles, we created a warning and awareness-building message as follows:
\begin{quotation}
Your way of playing may involve you in trouble. For your peace of mind and safety, let  us confirm the promises of the Pigg Party.
NOTE: This message is sent to those with similar usage trends as those involved in trouble in the past, aiming to mitigate trouble.

Pigg Party's Promises: \url{https://lp.pigg-party.com/rules}.
1) Do not disclose or ask for contact information, addresses, and locations of schools/workplaces.
2) Refrain from making promises to meet in person.
3) Do not send your own images.

The analysis algorithm used to send this message was developed through collaborative research conducted by Professor X of Z University and CyberAgent, Inc.
Currently, we operate on a trial basis to create an environment where everyone can use our services peacefully and safely.
\end{quotation}
This is an English translation, where X and Z are pseudonymized names \footnote{We will disclose the name of this university and professor after a review.}.
The academic researchers' names are disclosed because data utilization by academic institutions is highly socially acceptable~\citep{Morishita2021}.
The original message is shown in Fig.~\ref{fig_msg}.

.

\subsubsection{Randomized Controlled Trial}

We conducted a randomized controlled trial from February 13, 2022, to June 30, 2022.
We segregated all players into two groups: Intervention and control groups.

The risk-assessment system (Fig.~\ref{fig_ponchi}) described below assessed all players' violation risks and the risks of being violated. 
The system sent the top 100 players of each risk category in the intervention group. 
The players in the control group were not intervened by the system, regardless of their risk level.

Following these principles, we limited the number of players subjected to the intervention to a maximum of 200 per day. This comprised 100 players at a high risk of committing violations and 100 at a high risk of being violated. Note that the total number of players may be less than 200, as some players might be at a high risk of committing and being subjected to violations.
Similarly, we have a control group of high-risk players capped at 200 for comparison with the players who have been intervened.

No significant differences were observed between the intervention and control groups regarding demographic information, usage days, and the number of interventions.
Therefore, we could compare both groups without controlling for the covariates.
Table~\ref{tbl_cov} presents the basic statistics of these groups in the intervention experiments.
The number of listed cumulative high-risk players and the unique counts of high-risk players in the intervention and control groups were approximately the same.
The women and men ratio of high-risk players (64\% for unique players) was similar to that of all players (61\%).
The gender ratio difference between the groups was not significant at a significance level of 0.05, with a p-value of 0.343 on Pearson's chi-square test.
There were no significant differences at a significance level of 0.05 in age, usage days after interventions, the number of players who received a single intervention, and the number of players who received multiple interventions (five times or more).

The insignificance of usage days after interventions means that a negative impact of the intervention on the number of usage days was not detected.
This is essential for application providers because the number of players' usage days is a key performance indicator (KPI) of usage satisfaction.

Women were likelier to be one-time intervention players, whereas males received multiple interventions.
The risk assessment system determined that men were at higher risk of violation and being violated than women.

 \begin{table*}[t!]
  \begin{center}
  \caption{Basic statistics of the randomized control trial.
The cumulative total and unique players row exhibits the number of listed cumulative total high-risk players and unique counts of high-risk players, respectively, at each gender and group.
  Avatar ages and usage days after interventions indicate mean values and standard deviations for each gender and group (values in brackets are standard deviations).
  Usage days after interventions indicate the number of DM or AC usage days after 14 days from the next day of the interventions.
The student's t-test tested the age and usage days, and the number of intervention days (one time, five times or more) was tested using Pearson's chi-square test.  
}
    \label{tbl_cov}
         \tiny
    \begin{tabular}{l|rrr|rrr}
    \toprule 
    Avatar Gender & \multicolumn{3}{|c|}{Male} & \multicolumn{3}{|c}{Female} \\ \midrule
    Group	& Intervention & Control & p-value &  Intervention & Control & p-value \\  \midrule
    Cumulative Total & 8651 & 8531 && 12713 & 12873 &\\
    Unique Players & 4570 & 4497 && 8272 & 8347 &\\
    Avatar Age & 24.646 ($\pm 9.932$) & 24.325 ($\pm 9.678$) & 0.213 & 23.167 ($\pm 10.435$) & 23.141 ($\pm 10.455$) & 0.907\\
    Usage Days after Interventions & 5.761 ($\pm 3.736$) & 5.728 ($\pm 3.707$) & 0.604 & 5.445 ($\pm 3.762$) & 5.524 ($\pm 3.757$) & 0.141 \\
    1 time & 62.0\% & 61.6\% & 0.715& 75.0\% & 74.7\% & 0.733\\ 
    5 times or more & 7.11\% & 7.65\% & 0.347& 4.15\% & 3.92\% &0.477\\ 
    \bottomrule
\end{tabular}
\end{center}
\end{table*}

\subsection{Risk Assessment System}

We used a machine learning model, which assesses violation risk utilizing a dataset of the same platform (Pigg Party)~\citep{Nishiguchi2024}.
This model predicts violation probabilities for the next day with an AUC of 0.87
It uses multiple types of players' social networks and social behavior over the past two weeks as features.
The model also predicts the violation probability for each relationship.
Using this probability, we can estimate each player's risk of violation and being violated.

The model is divided into two phases: the weak learning phase, in which learning is based on a single network, and the stacking phase, in which the outputs of the weak learners are combined to construct the final model.
The weak learners are constructed based on Graph Attention Networks (GAT)~\citep{Velickovic2018}, one of the representative graph neural network architectures. The metadata and contact structures can be incorporated into the model by representing the player's metadata as nodes and the contacts between players as edges.

The metadata used included avatar age, avatar gender, number of friends, days since installation, and the activity levels of various functions, such as comments.
The gender and age of the avatar were set in the avatar and were optional inputs. 
Additionally, activity levels were divided by the number of login days for each player to eliminate biases owing to the number of login days.

The network defines the following five networks based on the five social interactions in Pigg Party. 
Although we focused on two main types of communication (AC and DM) in a randomized controlled trial on Pigg Party, we used all types of communication data to capture even the slightest signs for this model.
In the weak-learning phase, a GAT model was constructed for each network.
The five types of networks are as follows: 
\begin{itemize}
    \item {\bf AC network}: Player who spoke -- owner of room where they spoke,
    \item {\bf DM network}: Player who sent a DM -- player who received a DM,
    \item {\bf Comment network}: Player who commented -- player who was commented,
    \item {\bf Follow network}: Player who followed -- player who was followed,
    \item {\bf Like network}: Player who reacted -- player who received.
\end{itemize}


In the stacking phase, a method called stacking~\citep{Wolpert1992} was employed to construct a new model (metamodel) using the outputs of the GAT models as features. To train the metamodel, we applied LightGBM~\citep{Guolin2017}. LightGBM is an ensemble learning algorithm based on gradient boosting that can build high-performance models for tabular data.

The ground-truth labels for the model training were assigned based on whether the player engaged in violation relating to sexual grooming within the week immediately preceding the inference date. 
These ground-truth labels were assigned by the moderators of the Pigg Party application by their terms of service during their monitoring activities. This monitoring has been continuously conducted since the release of the Pigg Party application.
The violation risk for each player was assessed using the inference results of the metamodel. 

Here, we assumed that sexual predators who are not found through the administrator's monitoring also exhibit similar behavior.
This is because it is challenging to conceal the features of child grooming from their social behavior, although they can avoid posting direct sexual expressions.
We can expect that this model, using social behavior, could evaluate the risk of potential predators, even if they artfully slip through the monitoring process.

The risk assessment system used in this study extracted high-risk players for violations in each group as follows:
\begin{enumerate}
    \item The system excluded players listed as high-risk players in the past nine days.
    The reason for excluding players listed in the past nine days is that maintaining appropriate intervals is an effective strategy for the continuity of intervention effects~\citep{ito2018}.
    \item The system excluded players who received a penalty from the Pigg Party administrator, i.e., players the administrator has already discovered were excluded. 
    \item The system extracted players who have logged into Pigg Party for at least three days in the past seven days.
    This was based on the median activity level of the entire user base, calculated from the period immediately preceding the experiment (January 25-31, 2022).
    \item The system ranked them by the summations of the probability larger than 0.95.
    We call this summation value a risk score.
    Setting a high threshold of 0.95 for the risk value is prioritizing individuals with even one high-risk relationship over those with multiple small-risk relationships under the constraint of targeting a small number of players. 
    \item The system listed the top-100 rank players as high-risk players.
\end{enumerate}
The system listed high-risk players as being at risk of being violated for the intervention and control groups.
We refer to these listing days as the high-risk players' intervention days.

The high threshold of 0.95 is the 99.9th percentile (the median risk score was 0.32, and the 95th percentile score was 0.85).
After a thorough discussion with the Pigg Party provider, we decided to adopt this threshold of 0.95 for the following reasons: To minimize the negative impact of interventions, the Pigg Party provider and we considered target only users who pose a very high risk.

\begin{figure}[t!]
  \begin{center}
    \includegraphics[width=0.9\columnwidth]{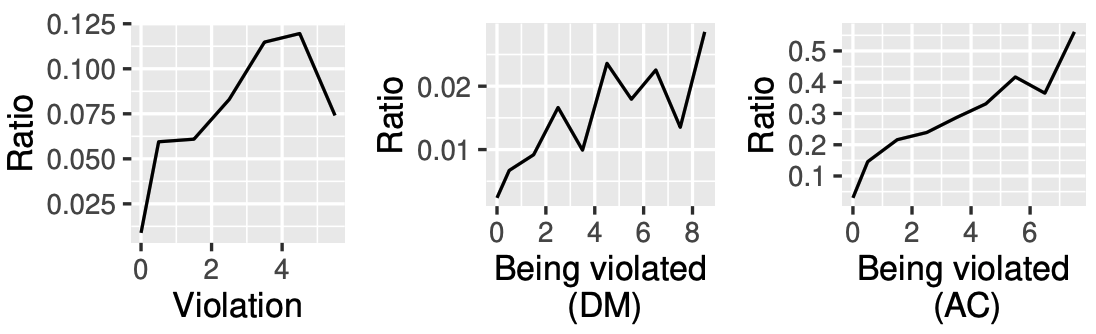}
    \caption{
The relationships between violation risk score and violation (left), the relationships between the risk score of being violated and being violated in DM (middle), and the relationships between the risk score of being violated and being violated in AC (right).}
    \label{fig_pred}
  \end{center}
\end{figure}

The risk scores for violation (or being violated) showed positive correlations with actual violations (or being violated) over the next 14 days (Fig.~\ref{fig_pred}).
As mentioned above, we assumed that detected violations and instances of being violated by the administrator do not always correspond to sexual predation and victimization. However, there should be a correlation between the risk score and actual violations or being violated.
Therefore, the positive correlation in this figure indicates the validity of this system.

This experiment had several variations, such as machine learning models and text contents of intervention messages.
We used only the above message and model~\citep{Nishiguchi2024} because our experiment involved actual application users, and the Pigg Party provider wanted to minimize intervention to reduce business risks. 
When we compare with additional messages and models, the intervention targets would have been increased, which could potentially have adverse effects. After carefully communicating with the service provider, we used only the above message and the model developed in \citep{Nishiguchi2024} for this study.

\subsection{Ethics}
The Pigg Party application provider collects behavior log data based on terms of service\footnote{\url{https://lp.pigg-party.com/terms}} and privacy policy\footnote{\url{https://www.cyberagent.co.jp/way/security/privacy/}}.
The provider explicitly stated the usage purpose and scope of the log data provision in these documents as follows: a) We use it for research purposes to develop new features and prevent violations; b) We aggregate the usage history generated in the course of providing our services and use it as primary data for academic research.
When users start a conversation for the first time, this service notifies them that the Pigg Party application provider will monitor their conversations to maintain a safe community. 
The dataset is pseudonymized.

The experiment analyzed in this study was carefully designed through meticulous meetings between the Pigg Party provider and the authors, in accordance with the terms of service, to ensure that players were not disadvantaged. 
The experiment did not disadvantage players, such as removing players' posts and banning players' accounts.
The analysis of this experimental data has received a non-applicability judgment from the ethics committee of a national university in Japan, considering that it has no impact on Pigg Party players\footnote{We will disclose the name of this university and the ethics review number after the review.}.

\section{Results}

\subsection{Inhibitation Effects of Violations (H1, H4)}

\begin{figure}[t]
  \begin{center}
    \includegraphics[width=0.5\columnwidth]{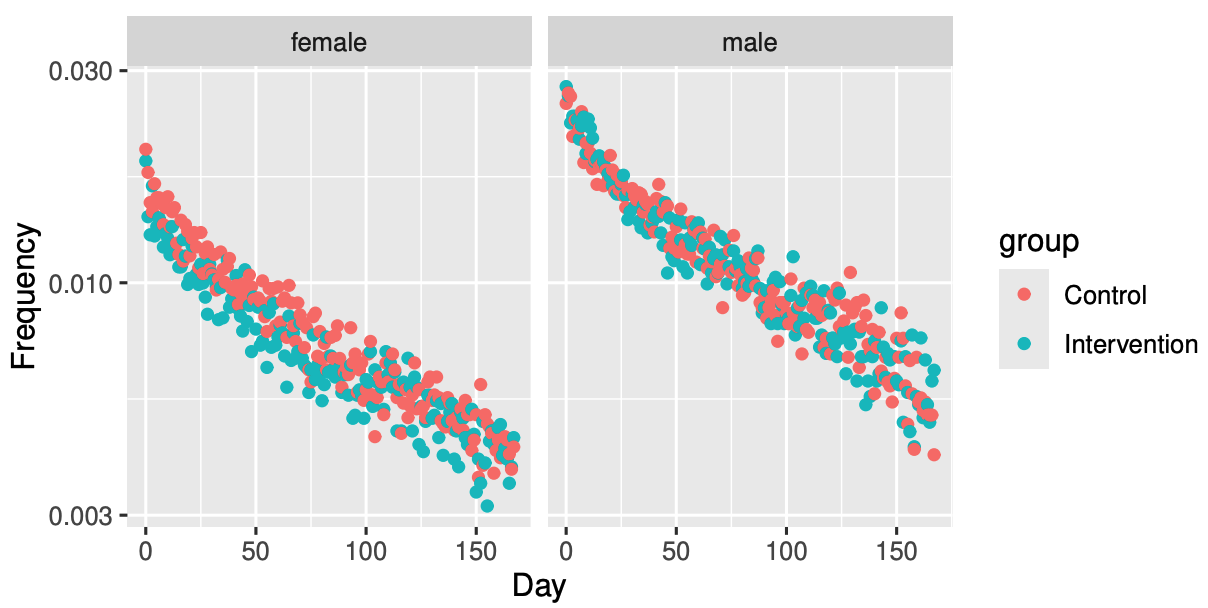}
    \caption{Violation frequencies of high-risk players and their time course from intervention days in the intervention and control groups.
    The vertical axes are log-scale.
    }
    \label{fig_day_eff}
  \end{center}
\end{figure}

Fig.~\ref{fig_day_eff} shows the violation frequencies of high-risk players and their time courses from the intervention days in the intervention and control groups.
Notably, we did not correspond potential predators/victims to violators/the being violated users in our data analyses.
The violation frequencies exponentially decreased with time.
Players' violation probabilities peaked when the risk assessment system listed them. 
The female violation frequencies showed the inhibitory effects of the intervention, whereas the male violation frequencies did not indicate such a significant difference.

Table~\ref{tbl_inhibit_ristex} indicates the results of the Fisher's exact test.
This intervention inhibited female violations over 12 weeks.

Repeated messaging diminished the effects of the intervention (Table~\ref{tbl_inhibit_ristex_n2}).
The intervention inhibited violations by women who received the message twice for four weeks.
The intervention did not significantly affect the violation frequency of the players who received the message thrice.

 \begin{table*}[t!]
  \begin{center}
  \caption{The inhibitory effects of the intervention on the violation of the high-risk players.
The effects are $(x_c-x_i)/x_c$, where the violation rates of the high-risk players in the intervention and control groups are $x_i$ and $x_c$, respectively.
Positive values indicate that an intervention decreased the violation rate compared with the control group.
The p-values were based on Fisher's exact test. This table shows the test results on seven periods for evaluating effect changes through days. 
We conducted the tests every two weeks in the first four weeks and for four weeks in subsequent weeks.
The effect values with p-values that are significant at a significance level of 0.05 are indicated by boldface type in the effect column.
The following tables also indicate the same presentation.
}
    \label{tbl_inhibit_ristex}
         \footnotesize
    \begin{tabular}{ll|rrrrrrr}
    \toprule 
    Gender	&	Value	&	1--14 days	&	15--28 days	&	29--56 days	&	57--84 days	&	85--112 days	&	113--140 days	&	141--168 days	\\	\midrule
Women 	&	Effect	&	{\bf 0.2157}	&	{\bf 0.1583}	&	{\bf 0.1117}	&	{\bf 0.1438}	&	0.0520	&	0.0814	&	0.0568	\\	
	&	p-value	&	0.0003	&	0.0207	&	0.0321	&	0.0061	&	0.3699	&	0.1493	&	0.3634	\\	\midrule
Men	&	Effect	&	0.1084	&	-0.0838	&	-0.0052	&	0.0184	&	-0.0990	&	0.0416	&	0.0422	\\	
	&	p-value	&	0.0907	&	0.2283	&	0.9323	&	0.7301	&	0.0808	&	0.4528	&	0.4731	\\	\bottomrule
\end{tabular}
\end{center}
\end{table*}

\begin{table}[t]
  \begin{center}
  \caption{The repeated messaging effects of the inhibition effects of the intervention on the violation of the high-risk players.
The top table shows players who received messages twice (top), and the bottom indicates players who received messages thrice (bottom).
  Tables~\ref{tbl_avoid_talk_ristex_n2} and \ref{tbl_avoid_chat_ristex_n2} also indicate the same presentation.
}\end{center}

    \label{tbl_inhibit_ristex_n2}
         \small
    \begin{tabular}{ll|rrr}
    \toprule 
    Gender	&	Value	&	1--14 days	&	15--28 days	&	29--56 days	\\	\midrule
Women	&	Effect	&	{\bf 0.3155}	&	{\bf 0.3670}	&	-0.0913\\
	&	p-value	&	0.0130	&	0.0097	&	0.5129 \\	\midrule
Men	&	Effect	&	0.0739	&	-0.1422	&	-0.0770\\
	&	p-value	&	0.6728	&	0.5128	&	0.6385\\ 
\bottomrule
    \toprule 
Women	&	Effect	&	-0.1319	&	-0.3237	&	-0.0279	\\
	&	p-value	&	0.6393	&	0.2469	&	0.8752	\\\midrule
Men	&	Effect	&	0.0806	&	0.0723	&	-0.1368	\\
	&	p-value	&	0.6849	&	0.6960	&	0.3900	\\
\bottomrule
\end{tabular}
\end{table}

\subsection{Effects of Avoidance of Being Violated (H2, H5)}

\begin{figure}[t!]
  \begin{center}
    \includegraphics[width=0.5\columnwidth]{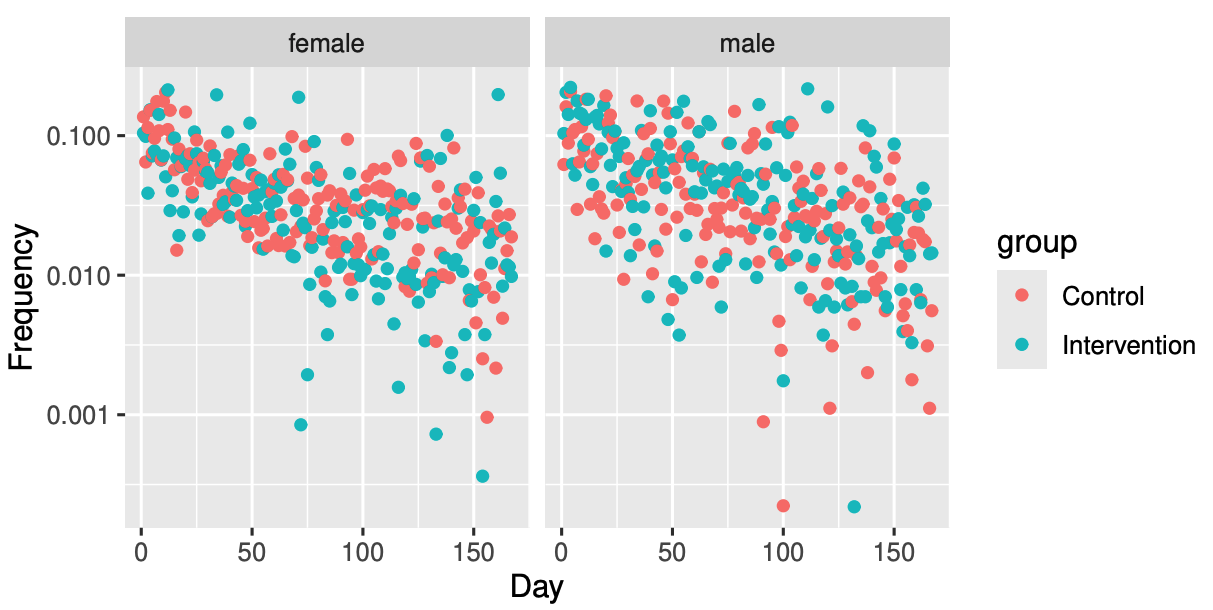}
    \includegraphics[width=0.5\columnwidth]{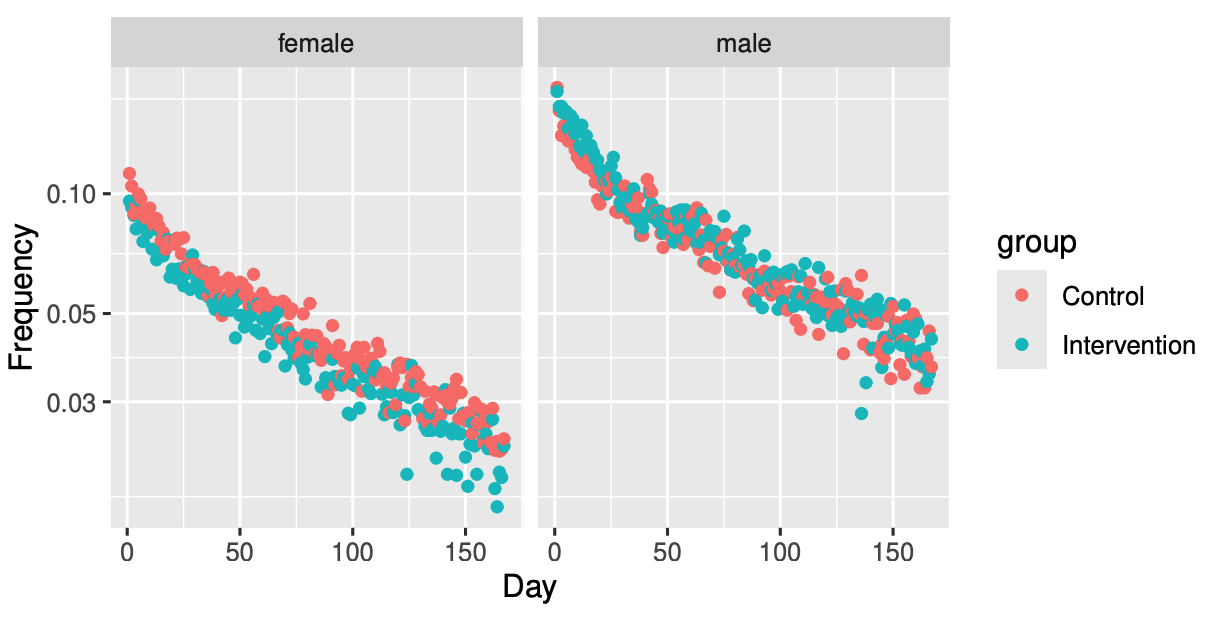}
    \caption{
The frequencies of instances of being violated among high-risk players and their time course from intervention days are shown for both the intervention and control groups (top: DM; bottom: AC).
    The vertical axes are log-scale.
    }
    \label{fig_day_eff_vic}
  \end{center}
\end{figure}

Fig.~\ref{fig_day_eff_vic} shows the frequencies of instances of being violated among high-risk players and their time course from the intervention days in both the intervention and control groups.
These frequencies exponentially decreased with time.
The frequencies of women being violated in the AC showed inhibitory effects of the intervention, whereas the frequency for males and those in the DM did not show such a difference.

Tables ~\ref{tbl_avoid_talk_ristex} and \ref{tbl_avoid_chat_ristex} indicate the results of Fisher's exact tests for DM and AC, respectively.
In DM, female players avoided being violated in the first two weeks.
They avoided violations in the AC for over 12 weeks, excluding 29–56 days.

The intervention for the players who received repeated messaging did not show any significant effects (Tables ~\ref{tbl_avoid_talk_ristex_n2} and \ref{tbl_avoid_chat_ristex_n2}).

 \begin{table*}[t]
  \begin{center}
  \caption{
The avoidance effects of the intervention on the instances of being violated in DM of the high-risk players.
}\end{center}

    \label{tbl_avoid_talk_ristex}
         \footnotesize
    \begin{tabular}{ll|rrrrrrr}
    \toprule     
    Gender	&	Value	&	1--14 days	&	15--28 days	&	29--56 days	&	57--84 days	&	85--112 days	&	113--140 days	&	141--168 days	\\	\midrule
Women &	Effect	&	{\bf 0.2646}	&	0.1725	&	0.0377	&	0.1794	&	0.0752	&	0.0971	&	0.1081	\\	
	&	p-value	&	0.0257	&	0.2189	&	0.7601	&	0.1287	&	0.5779	&	0.4761	&	0.4749	\\	\midrule
Men	&	Effect	&	-0.0491	&	-0.0094	&	0.1046	&	0.0288	&	-0.0145	&	0.1994	&	-0.0740	\\	
	&	p-value	&	0.8005	&	1.0000	&	0.4675	&	0.8820	&	0.9400	&	0.1863	&	0.7197	\\	\bottomrule
\end{tabular}
\end{table*}

\begin{table*}[t]
  \begin{center}
  \caption{
The avoidance effects of the intervention on the instances of being violated in AC of the high-risk players
}
\end{center}
     \label{tbl_avoid_chat_ristex}
     \footnotesize
    \begin{tabular}{ll|rrrrrrr}
    \toprule 
        Gender	&	Value	&	1--14 days	&	15--28 days	&	29--56 days	&	57--84 days	&	85--112 days	&	113--140 days	&	141--168 days	\\	\midrule
Women &	Effect	&	{\bf 0.1111}	&	{\bf 0.1105}	&	0.0414	&	{\bf 0.0837}	&	0.0594	&	0.0648	&	{\bf 0.0738}	\\	
	&	p-value	&	0.0023	&	0.0058	&	0.1715	&	0.0063	&	0.0694	&	0.0525	&	0.0311	\\	\midrule
Men	&	Effect	&	-0.0059	&	-0.0451	&	0.0096	&	0.0248	&	-0.0389	&	0.0082	&	-0.0126	\\	
	&	p-value	&	0.8866	&	0.3056	&	0.7585	&	0.4073	&	0.2270	&	0.8105	&	0.7233	\\	\bottomrule
\end{tabular}
\end{table*}

 \begin{table}[t]
  \begin{center}
  \caption{
The repeated messaging effects of the intervention's avoidance effects on the instances of being violated in DM among the high-risk players.
}\end{center}

    \label{tbl_avoid_talk_ristex_n2}
         \small
    \begin{tabular}{ll|rrr}
    \toprule 
    Gender	&	Value	&	1--14 days	&	15--28 days	&	29--56 days	\\	\midrule
Women	&	Effect	&	-0.0286	&	0.1567	&	0.1423	\\
	&	p-value	&	0.9241	&	0.4546	&	0.4423	\\	\midrule
Men	&	Effect	&	-0.0563	&	0.1741	&	-0.0563	\\
	&	p-value	&	0.8961	&	0.6385	&	0.8961	\\
\bottomrule
    \toprule 
Women &	Effect	&	0.2057	&	0.1078	&	0.0183	\\
	&	p-value	&	0.4205	&	0.7576	&	1.0000	\\	\midrule
Men	&	Effect	&	0.3360	&	0.0523	&	0.0523	\\
	&	p-value	&	0.2941	&	1.0000	&	1.0000	\\
\bottomrule
\end{tabular}
\end{table}

 \begin{table}[t]
  \begin{center}
  \caption{The repeated messaging effects of the intervention's avoidance effects on the instances of being violated in AC among the high-risk players.
}
 \end{center}
   \label{tbl_avoid_chat_ristex_n2}    
         \small
\begin{tabular}{ll|rrr}
    \toprule 
    Gender	&	Value	&	1--14 days	&	15--28 days	&	29--56 days	\\	\midrule
Women	&	Effect	&	0.0614	&	0.1290	&	0.0582	\\
	&	p-value	&	0.1997	&	0.0151	&	0.2000	\\\midrule
Men	&	Effect	&	0.0901	&	-0.0492	&	0.0665	\\
	&	p-value	&	0.1691	&	0.5664	&	0.2975	\\
\bottomrule
    \toprule 
Women	&	Effect	&	0.0972	&	0.0422	&	0.0840 \\
	&	p-value	&	0.1581	&	0.5547	&	0.0795\\	\midrule
Men	&	Effect	&	0.0075	&	-0.0411	&	0.0165\\
	&	p-value	&	0.9509	&	0.5873	&	0.7861\\
\bottomrule
\end{tabular}
\end{table}

\subsection{Behavior Change (H3)}

Table~\ref{tbl_midnight_usage} shows each group's DM and AC usage from 8 P.M. to 5 A.M. during the 84 days after the intervention because violations occur at these times in Pigg Party~\citep{Yokotani2021_chb2}.
Female players in the intervention group, who experienced fewer being violated because of the intervention, reduced their usage at night compared to the control group.
In contrast, male players did not exhibit a similar reduction in nighttime usage.

\begin{table}[t]
  \begin{center}
  \caption{
  The change of night usage for each gender.
 As ``night usage,'' we counted the number of windows each player used DM or AC from 8 P.M.  to 5 A.M., considering hourly windows, i.e., the maximum value is 93.
 This shows the counted value is an average of 84 days for each player.
 We compared the intervention and control groups with the Wilcoxon rank sum test (p-values: 0.0005 (women) and 0.4309 (men)).
Women in the intervention group experienced reduced night usage at a significance level of 0.05.
}
  \end{center}
   \label{tbl_midnight_usage}
     \small
    \begin{tabular}{ll|rrrrr}
    \toprule 
    Gender & Group & 2.5\%ile & 25\%ile & 50\%ile & 75\%ile & 97.5\%ile\\\midrule   
    Women & Intervention & 0.000 & 0.024 & 0.107 & 0.310 & 0.845\\
           & Control      & 0.000 & 0.024 & 0.119 & 0.321 & 0.881\\ \midrule           
    Men & Intervention & 0.000 & 0.048 & 0.167 & 0.381 & 0.845 \\
           & Control      & 0.000 &  0.048 & 0.179 & 0.381 & 0.857 \\ \bottomrule
\end{tabular}
\end{table}

\section{Discussion}

We established an intervention method to inhibit sexual grooming based on the routine activity theory~\citep{Lawrence1979} in criminal psychology, developed an intervention system as part of a risk assessment system, and quantitatively indicated its effectiveness in a randomized control trial on an actual online communication platform.

This messaging intervention inhibited high-risk female players' violations and risky behavior related to being violated on the AC.
This is because they reduced risky behavior, such as nighttime usage, which resulted in violations~\citep{Yokotani2021_chb2,Parapar2014,Rodrigues2016}.
However, this intervention did not impact men.
Thus, {\bf H1}, {\bf H2}, and {\bf H3} were supported by female players.
Women accepted and responded to the warnings and awareness-building messages related to sexual predation.
This may be because they are more often victims of sexual perpetration compared to men~\citep{Mohler-Kuo2014,Henry2016,DeKeseredy2021}.
This is an essential result in terms of reducing online risk for those at a high risk of sexual victimization.

Although the intervention was simple and inexpensive, the impact was long-lasting (12 weeks).
{\bf H4} and {\bf H5} were supported, but the impact was surprisingly long-lasting.
This implies that simple interventions such as warnings and awareness-building messaging can be effective, similar to findings in previous studies on racial discrimination posts~\citep{Munger2017} and energy conservation~\citep{ito2018}.

Interventions to avoid being violated had long-term effectiveness in AC but did not last in DM. 
This might be because, compared to AC, where it is easy to prevent dangerous players by moving rooms, DM groups, formed based on friend connections, have low fluidity of relationships, making it challenging to avoid dangerous players. 
This suggests that even with awareness building, changing established relationships with dangerous individuals is challenging if relational fluidity is low.

The inhibition effects on the violations and instances of being violated weakened for players who received repeated messaging.
Although the female players who received the messages twice temporarily reduced violations, the intervention did not show significant effects under other conditions.
This could be owing to habituation~\citep{Thompson1966,Groves1970,Rankin2009} or persistent players, less likely to be influenced by the intervention, receiving the messages repeatedly.
However, it was challenging to distinguish between them in this study.
The effects of repeated intervention require further investigation to understand their implications and potential applications fully.

The intervention did not affect males.
This might be because the gender of sexual predators tends to be men than women~\citep{Winters2017}.
The intervention messages were indirect warnings, as the message targets were both potential victims and predators.
This suggests that the presence of monitoring by the administrator through warning messages did not inhibit their violation behavior.
Additionally, this result may indicate that the intervention did not support potential male victims.
Male-child-victims are not a small number~\citep{Mohler-Kuo2014,Kaylor2022}, and sexual abuse against them has long-term adverse psychological effects~\citep{Spataro2004,Dube2005}.
Therefore, it is essential to develop interventions to increase awareness among men.

The intervention did not negatively affect the players' continuous usage of irrelevant gender.
This approach maintains user satisfaction (application providers' KPI).
This was essential to implementing the intervention.

Our findings provide advanced evidence for routine activity theory~\citep{Lawrence1979} in criminal psychology.
Previous studies on cyber offenders' routine online activities are mostly qualitative~\citep{Santisteban2018}.
There is little evidence of online user behavior, as \citep{Yokotani2021_chb2}.
Our findings provide novel evidence for the effectiveness of warnings and awareness-building interventions based on a randomized controlled trial involving more than 24,000 actual online communication platform users.
The intervention by sending warnings and awareness-building messages can inhibit risky behavior of ``suitable targets;'' thus, they decreased the frequency of the violation and being violated.
This inhibition occurs because of the liquidity of social relationships.
On the other hand, ``a likely offender'' did not change their behavior owing to the intervention; thus, they did not decrease their frequency of the violation and being violated.
Exuding the presence of an application administrator as ``a capable guardian'' through indirect warning messages may not be compelling enough.

\subsection{Limitations and Future Works}

We tested the intervention effects based on violations extracted through monitoring by application administrators according to the terms of services.
As mentioned above, sexual predators artfully slip through the monitoring~\citep{Lykousas2018,Lykousas2021,Ringenberg2022,Borj2023}.
Evaluations using indicators that more accurately represent actual victimizations are required, such as reports of victimization by players.

We did not distinguish potential predators and victims due to the restriction of the dataset.
This is because violators or users who are being violated do not always correspond to predators or victims.
To distinguish them, language information on sexual grooming communication seems to enable estimate high-risk users to predators or victims.
However, utilizing it is strictly restricted by Japan's constitutionally guaranteed secrecy of communication.
If obtaining valid consent for analyzing the data, it can estimate high-risk users to be predators or victims.
This would allow us to evaluate the effects of our intervention on predators and victims, and we would understand their behavior better.

This study highlights the limitations of this intervention, which did not affect men. Developing interventions to raise awareness among males is essential to inhibit male predators and support male victims.
The limited effect of the intervention on reducing risky behavior among male users might be attributed to their difficulty in envisioning themselves as either perpetrators or victims. 
To help them know the victimization risks, an approach that raises awareness about male sexual violence victims could be practical~\cite{Kaylor2022,Mohler-Kuo2014}. 
Regarding the perpetrator image, it is essential to highlight that in Japan, luring minors is a crime even with the minor's consent, and it remains a crime even without any sexual contact. However, several people seem to be unaware of this. Therefore, educating male participants about the apparent criminality of luring minors could be beneficial. 

We can consider several additional settings in our experiment, such as other machine learning models, other message contents, and parameter settings.
The content of the messages can indeed impact the degree of behavior change, as evidenced by a behavioral economics study~\cite{ito2018} on energy-saving messages. 
We acknowledge this as necessary for further investigation.

The study was conducted using a Japanese avatar communication application.
Examining the proposed intervention on other platforms would provide insights into more robust and effective interventions.

\section{Conclusion}

We proposed an intervention method for inhibiting sexual grooming (perpetration and victimization). This was based on criminal psychology, which developed a risk assessment system for the intervention and evaluated the intervention effects via a randomized controlled trial of warnings and awareness-building messaging for high-risk players. Consequently,  our proposed intervention inhibited females' online risks related to sexual abuse.
Our findings would contribute to the fight against online sexual abuse and advance criminal psychology.

\section*{Data Availability Statement}
The data cannot be shared publicly because of the privacy policy of CyberAgent, Inc. 
It prohibits publishing users' personal data. The data are shared by CyberAgent, Inc. for researchers within the scope of the informed consent obtained from subjects. The data that support the findings of this study are available upon reasonable request from the authors.

\bibliographystyle{linquiry2}


\section*{Competing Interests}

Masanori Takano is an employee of CyberAgent, Inc. 
There are no patents, products in development, or marketed products to declare. 
Fujio Toriumi was funded by CyberAgent, Inc.

\section*{Acknowledgement}
We are grateful to Dr. Soichiro Morishita at CyberAgent, Inc. whose comments and suggestions were very valuable throughout this study.

\clearpage
\section*{Supplementary Information}

\setcounter{equation}{0}
\setcounter{section}{0}
\setcounter{figure}{0}
\setcounter{table}{0}

\renewcommand{\thesection}{S\arabic{section}}

\renewcommand{\figurename}{Fig.}
\renewcommand{\thefigure}{S\arabic{figure}}
\renewcommand{\tablename}{Table.}
\renewcommand{\thetable}{S\arabic{table}}

\renewcommand{\theequation}{S\arabic{equation}}
\appendix

\begin{figure}[hb!]
  \begin{center}
    \includegraphics[width=0.5\columnwidth]{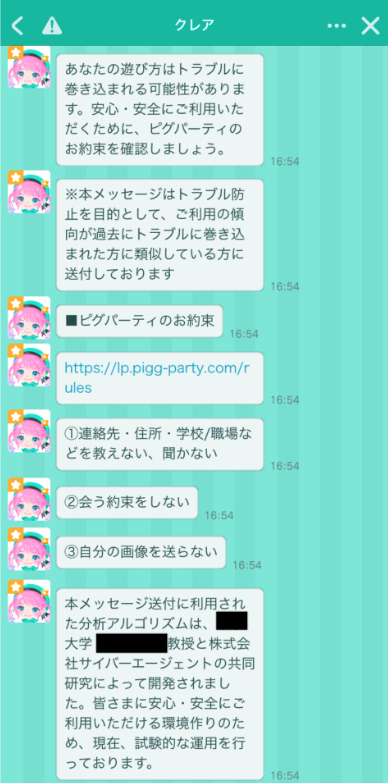}
    \caption{Warning and awareness building message from the Pigg Party administrator (original Japanese version)}
    \label{fig_msg}
  \end{center}
\end{figure}

\end{document}